\documentclass[aps,prl,reprint,amsmath,amssymb,twocolumn,superscriptaddress,notitlepage,groupedaddress]{revtex4-1}
\usepackage{graphicx}
\usepackage{dcolumn}
\usepackage{stackengine}
\usepackage[dvipsnames*,svgnames]{xcolor}
\usepackage{verbatim}
\usepackage[position=bottom,caption=false,captionskip=3pt,font=normalsize,subrefformat=simple,labelformat=simple]{subfig}

\usepackage{listings}
\usepackage{mathtools}
\usepackage{amsmath}
\usepackage{chngcntr}
\usepackage{bm}
\usepackage{url}

\usepackage[breaklinks]{hyperref}
\usepackage{breakurl}
\usepackage{tabularx}
\usepackage{soul}
\hypersetup{
colorlinks,
linkcolor={blue!100!black},
citecolor={blue},
urlcolor={blue!80!black}}
\def\equationautorefname~#1\null{\,\,(#1)\null}
\begin{document}
\title{Phonon Bottleneck Identification in Disordered Nanoporous Materials}
\author{Giuseppe Romano}
\email{romanog@mit.edu}
\affiliation{Department of Mechanical Engineering, Massachusetts Institute of Technology, 77 Massachusetts Avenue, Cambridge, MA 02139, USA}
\author{Jeffrey C. Grossman}
\affiliation{Department of Materials Science and Engineering, Massachusetts Institute of Technology, 77 Massachusetts Avenue, Cambridge, MA 02139}
\begin{abstract}
Nanoporous materials are a promising platform for thermoelectrics in that they offer high thermal conductivity tunability while preserving good electrical properties, a crucial requirement for high-efficiency thermal energy conversion. Understanding the impact of the pore arrangement on thermal transport is pivotal to engineering realistic materials, where pore disorder is unavoidable. Although there has been considerable progress in modeling thermal size effects in nanostructures, it has remained a challenge to screen such materials over a large phase space due to the slow simulation time required for accurate results. We use density functional theory in connection with the Boltzmann transport equation, to perform calculations of thermal conductivity in disordered porous materials. By leveraging graph theory and regressive analysis, we identify the set of pores representing the phonon bottleneck and obtain a descriptor for thermal transport, based on the sum of the pore-pore distances between such pores. This approach provides a simple tool to estimate phonon suppression in realistic porous materials for thermoelectric applications and enhance our understanding of heat transport in disordered materials.\end{abstract}
\maketitle

\section*{Introduction}
The efficient and inexpensive conversion of heat directly into electricity is a long-sought goal with enormous potential in the clean-energy technology landscape~\cite{CRC1995,tian2013heat}. The engineering of thermoelectric materials, however, is particularly challenging because of the interrelation of key physical properties constituting the thermoelectric figure of merit ZT, defined as $ZT = \frac{T\sigma S^2}{\kappa}$ where $\sigma$ is the electrical conductivity, $\kappa$ is the lattice thermal conductivity, $S$ is the Seebeck coefficient, and $T$ the temperature. Nanostructuring offers a powerful way to decouple the electrical and thermal transport. In most semiconductors, the numerator of ZT, also referred to as ``power factor,'' is maximized at relatively high carrier concentrations, so the dominant electron mean free path (MFP) can be as small as a few nanometers~\cite{liao2015significant}. Conversely, phonons may have much larger MFPs, even on the order of microns~\cite{esfarjani2011heat}. Properly engineered nanostructures are therefore able to scatter phonons more effectively than electrons. Porous materials offer a highly tunable platform thanks to their great degree of structural tunability including pore size, shape, and arrangement, as well as the potential for controllable uniform thin films, high temperature resilience and robust contacts. As an example, the thermal conductivities of nanoporous Si have been measured in many studies with the common finding of a strong suppression of thermal transport, leading to a significant improvement in experimentally measured ZT~\cite{Hopkins2011,song2004thermal,lee2015ballistic,Tang2010,Yu2010,verdier2017thermal,perez2016ultra,lee2017investigation}. On the computational level, several models based on the Boltzmann transport equation (BTE) also have shown low thermal conductivities and revealed significant features of phonon-boundary scattering and fundamental thermal transport in nanoporous materials~\cite{hao2009frequency,lee2008nanoporous,romano2015}. Preliminary attempts aiming at tuning thermal conductivity in nanoporous Si have shown that, even within ordered configurations and with pores of the same size, the pattern of the pores can have a large influence on the resulting thermal transport~\cite{Romano2014qj}. Although aligned configurations offer a robust platform for controllable experiments, pore disorder is unavoidable, especially at smaller length scales~\cite{smith2016catalyst}. Recent Monte Carlo calculations~\cite{wolf2014monte,wolf2014thermal} investigated thermal transport in disordered porous materials with circular pores and concluded that the density of pores along the heat flux direction has a significant influence on thermal conductivity. In this paper, we expand on this concept by developing a method that identifies the actual set of pores representing the highest local resistance to phonon transport. To this end, we use the recently developed first-principles BTE solver~\cite{romano2016temperature} to perform thermal transport calculations in random-pore configurations with pores of circular and square shapes. Then, we establish a correlation between the phonon suppression and the pore arrangement within a given configuration, leading to the identification of the pores constituting the phonon bottleneck. Upon introducing a simple descriptor representing the strength of this collection of pores, we find a correlation between such a parameter and the effective thermal conductivity $\kappa_{eff}$. This work can be potentially used to estimate the degree of phonon suppression in realistic nanoporous samples while avoiding the computational burden of solving the BTE.

\section*{Phonon Boltzmann Transport Equation}
Our computational approach is based on our recent implementation of the BTE for phonons, which under the relaxation time approximations, reads as ~\cite{romano2015} \begin{equation}
\begin{split}\label{Eq:1}
\Lambda \mathbf{\hat{s}}(\Omega) \cdot \nabla T(\mathbf{r},\Omega,\Lambda) + T(\mathbf{r},\Omega,\Lambda)=\\ \gamma \int \frac{K(\Lambda')}{\Lambda'^2}<T(\mathbf{r},\Omega',\Lambda')>d\Lambda', 
\end{split}
\end{equation}where $K(\Lambda)$ is the bulk MFP distribution, $T(\mathbf{r},\Omega,\Lambda)$ is the effective temperature associated to phonons with MFP $\Lambda$ and direction, $\mathbf{\hat{s}}(\Omega)$, denoted by the solid angle $\Omega$. The term $\gamma = \left[\int K(\Lambda)/\Lambda^2 d\Lambda\right]^{-1}$ is a bulk material property, and $<.>$ is an angular average. The RHS of Eq.\autoref{Eq:1} is the effective lattice temperature, a quantity describing the average phonon energy. The term $K(\Lambda)$ is obtained by using harmonic and anharmonic forces in connection with density functional theory~\cite{broido2007intrinsic,esfarjani2011heat}. The spatial discretization of Eq.\autoref{Eq:1} is achieved by the finite-volume (FV) method. The simulation domain is discretized by means of an unstructured mesh, generated by GMSH~\cite{geuzaine2009gmsh}. The phonon BTE requires the solid angle discretization to account for different phonon directions. We use the discrete ordinate method (DOM), a technique that solves the BTE for each phonon direction independently and then combines the solutions by an angular integration~\cite{abe1997derivation}. As Si is a nongray material, i.e., has a broad MFP distribution, we need to discretize the MFP space, as well. We reach convergence with 30 MFPs (uniformly distributed in log space) and 576 phonon directions. The algorithm is detailed in Ref.~\cite{romano2011multiscale}. The overall solution of Eq. 1 requires solving the BTE thousands of times, leading to an increase in the computational time. However, our solver has been conveniently parallelized and each configuration takes only a few minutes with a cluster of 32 nodes.

The walls of the pores are assumed diffusive, a condition that translates into \begin{equation}
\begin{split}\label{Eq:2}
T_b = - \frac{\int_{\Omega^+}\int \left(K(\Lambda)/\Lambda\right) T(\mathbf{r},\Omega,\Lambda)\mathbf{\hat{s}}(\Omega)\cdot\mathbf{\hat{n}}\,d\Omega d\Lambda}{\int_{\Omega^-}\int \left(K(\Lambda)/\Lambda\right) \mathbf{\hat{s}}(\Omega)\cdot\mathbf{\hat{n}} \, d\Omega d\Lambda}, 
\end{split}
\end{equation}where $\Omega^-$ and $\Omega^+$ are the solid angle for incoming and outgoing phonons with respect to the contact with normal $\mathbf{\hat{n}}$.  Once Eq.\autoref{Eq:1} is solved, thermal flux is computed via $\mathbf{J}(\mathbf{r}) = 3\int K(\Lambda)/\Lambda <T(\mathbf{r},\Omega,\Lambda)\mathbf{\hat{s}}(\Omega)> d\Lambda$. The effective thermal conductivity is obtained by using Fourier's law, i.e., $\kappa_{bte} = (L/\Delta T) \int_{hot} <\mathbf{J}(\mathbf{r},\Omega,\Lambda)\cdot\mathbf{\hat{n}}> dS $, where $\Delta T$ = 1 K is the applied temperature and K is the distance between the hot and cold contacts (or the size of the unit cell). To focus on phonon size effects, we normalize the thermal conductivity by its diffusive value, i.e., $\kappa_{eff} = \kappa_{bte}\left(\kappa_{bulk}/\kappa_{fourier}\right)$, where $\kappa_{fourier}$ is the thermal conductivity computed by the diffusive heat equation and $\kappa_{bulk}$ = 156 Wm$^{-1}$K$^{-1}$ is the bulk thermal conductivity~\cite{Li2014fg}. Our approach has been validated against experiments on porous silicon~\cite{song2004thermal,vega2016thermal} and, more recently, on silicon labyrinths~\cite{goodson}. 
\section*{Aligned and Random Pores}
\begin{figure*}

\subfloat[\label{Fig:10a}]
{\includegraphics[width=0.48\textwidth]{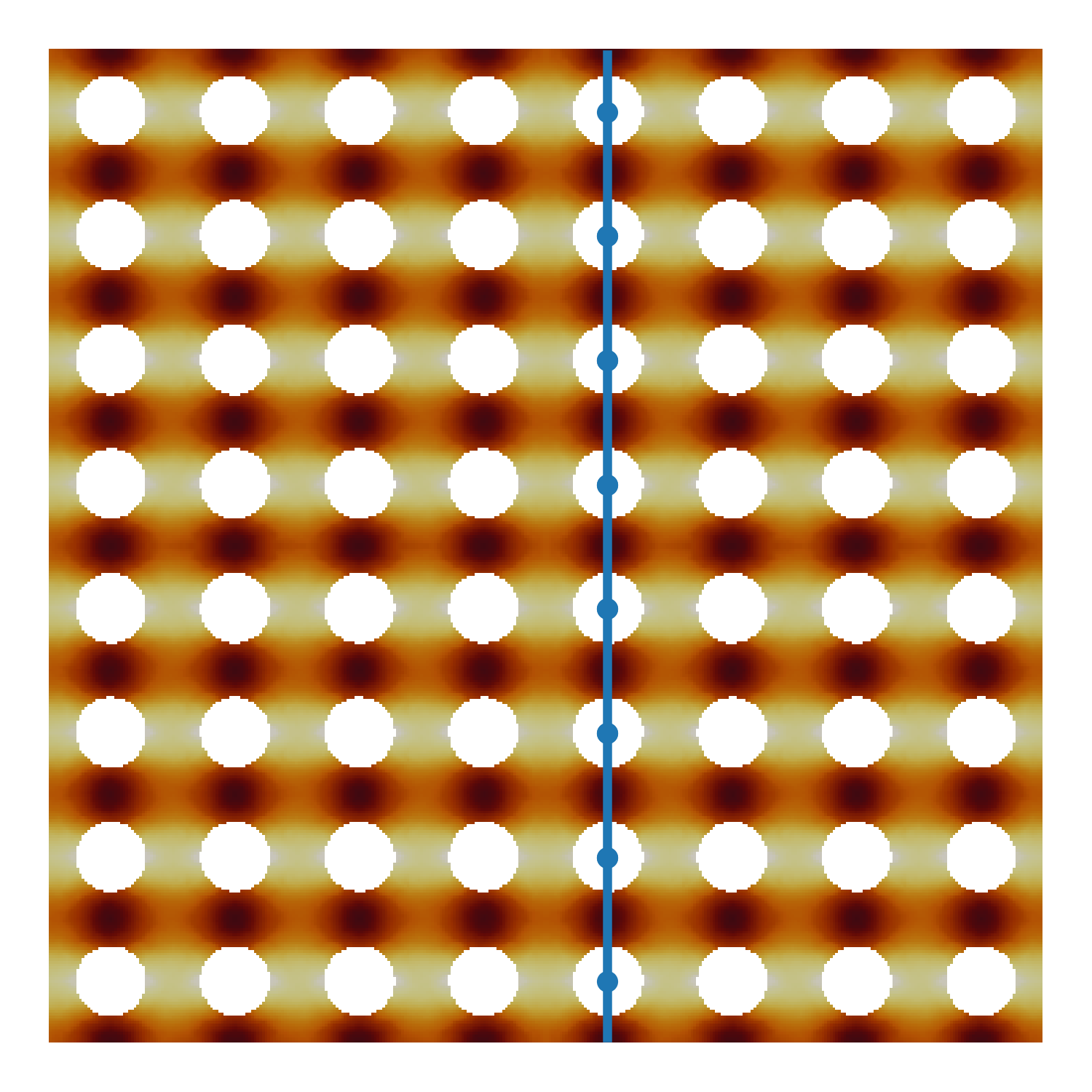}}
\hfill
\subfloat[\label{Fig:10b}]
{\includegraphics[width=0.48\textwidth]{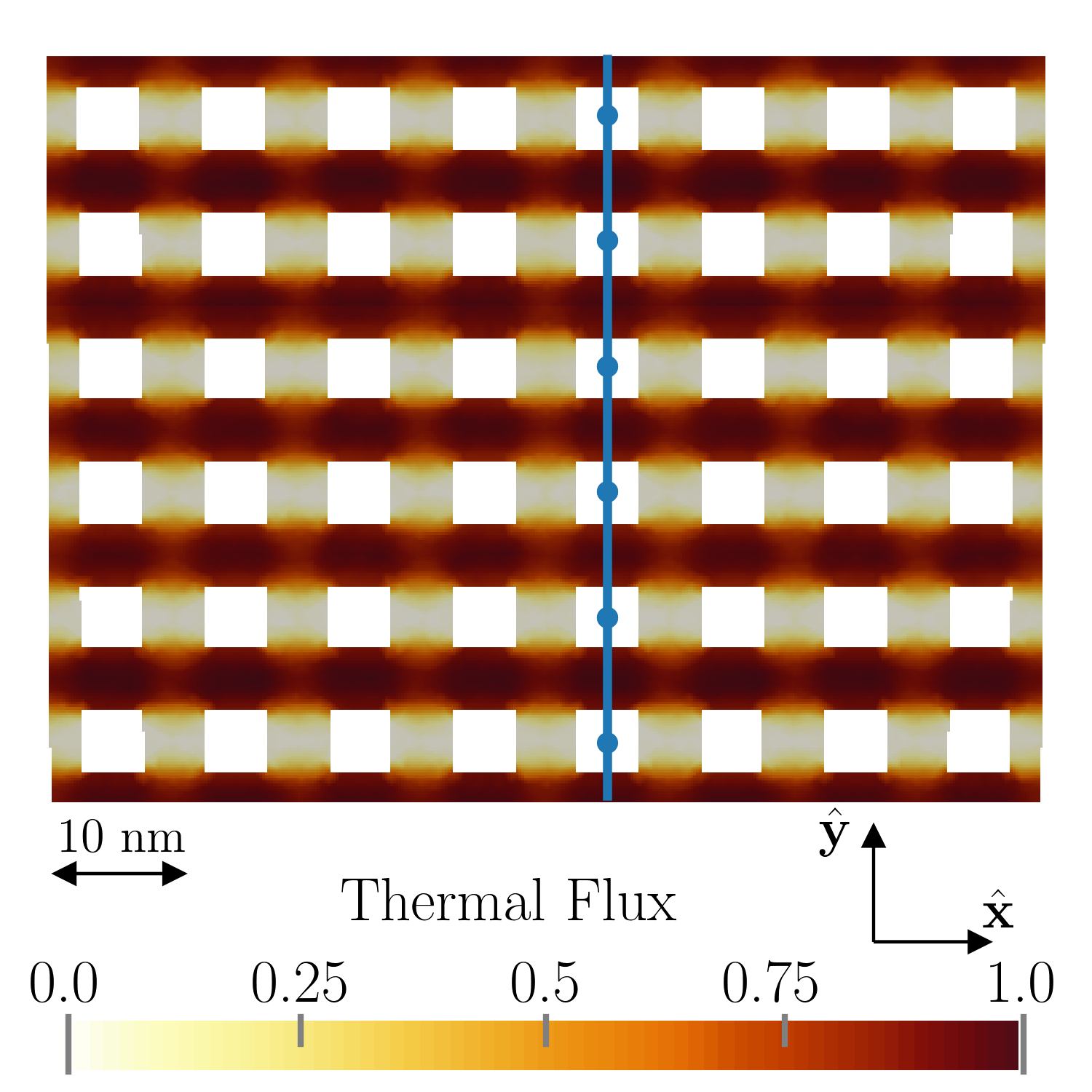}}
\quad
\subfloat[\label{Fig:10c}]
{\includegraphics[width=0.48\textwidth]{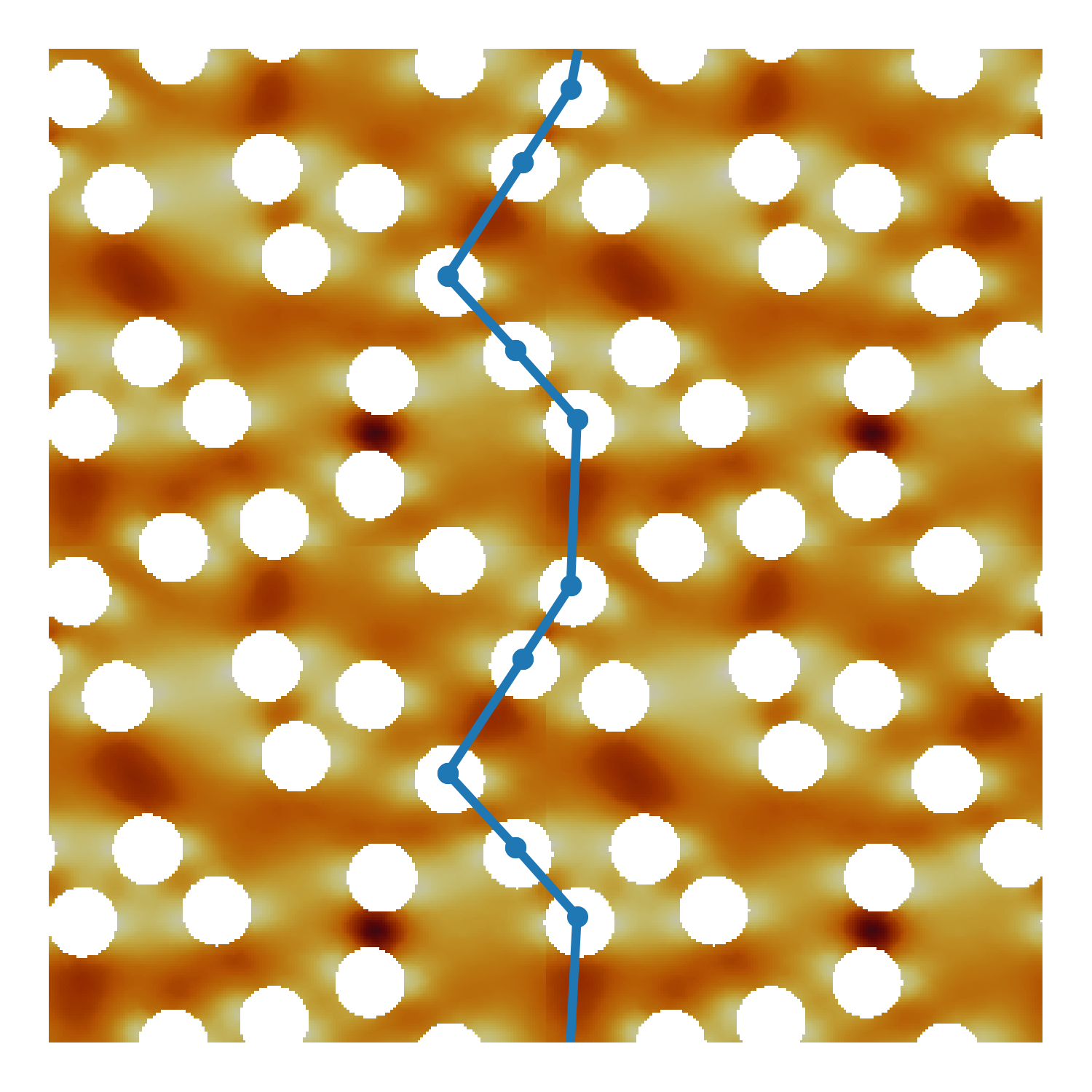}}
\hfill
\subfloat[\label{Fig:10d}]
{\includegraphics[width=0.48\textwidth]{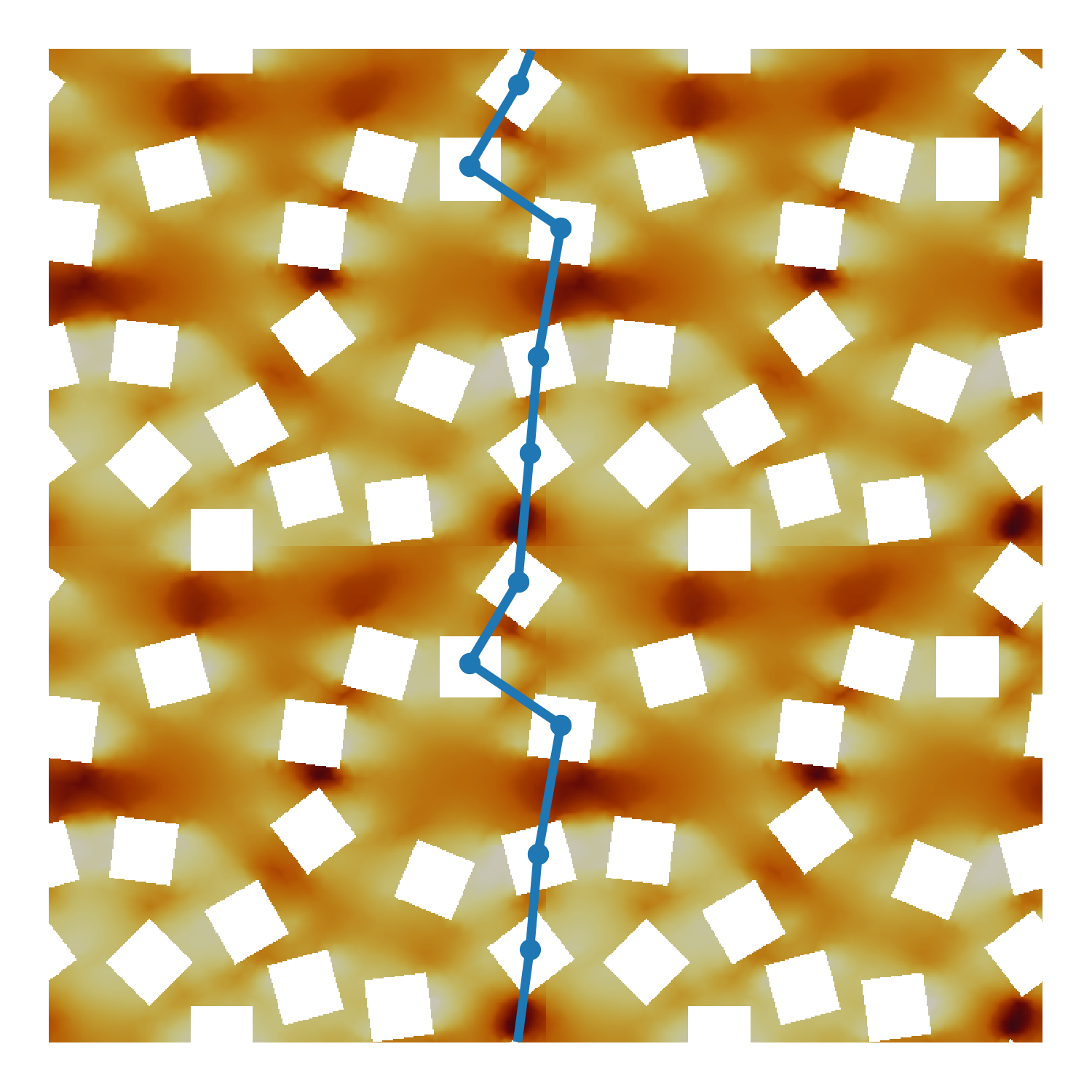}}
\caption{Normalized magnitude of thermal flux for the (a) AC, (b) AS, (c) DC and (d) DS cases. The temperature gradient is imposed along the $x$ direction. Phonons prefer to travel in the spaces between the pores, as highlighted by the red areas. In all the configurations the pores arrangement is periodic in both $x$ and $y$ directions. The blue line represents the phonon bottleneck.}
\end{figure*}

We first compute thermal transport in aligned configurations, which we will refer to as ``aligned circular'' (AC) and ``aligned square'' (AS). The unit cell comprises a single pore and is a square with size L = 10 nm. Heat flux is enforced by applying a difference of temperature $\Delta T$ = 1 K along the $x$ direction. The porosity is fixed at $\phi = 0.25$, and periodic boundary conditions are applied throughout. The computed values for $\kappa_{eff}$ are in both cases around 10 W m$^{-1}$ K $^{-1}$, considerably lower than $\kappa_{bulk}$. The magnitude of heat flux, shown in Fig.~\subref*{Fig:10a} and in Fig.~\subref*{Fig:10b} for AC and AS, respectively, indicates that phonon travel mostly near the spaces between pores perpendicular to the applied temperature gradient.

For random-pore (or disordered) configurations, the size of the unit cell is chosen to be L = 40 nm, four times as large as that for the aligned cases, in order to generate significant disorder in the pores arrangement. Sixteen nonoverlapping pores are randomly placed while keeping the porosity fixed to $\phi$ = 0.25, thus allowing a direct comparison with the aligned counterparts. We note that the material is still periodic in that pores crossing the border of the unit cell are repeated in the adjacent unit cells. We compute $\kappa_{eff}$ for two hundred arrangements, one hundred for each shape, which we refer to as ``disordered circle'' (DC) and ``disordered square'' (DS). The magnitude of thermal flux for two configurations is shown in Fig.~\subref*{Fig:10c} and Fig.~\subref*{Fig:10d}, respectively. We note that the formation of high-flux regions is irregular as it depends on the pore configuration. According to Fig.~\subref*{Fig:30a} and Fig.~\subref*{Fig:30b}, respectively, the DC and DS cases are found to have average $\kappa_{eff}$ values 15 $\%$ and 30 $\%$ lower than that of their aligned counterparts. Intuitively, the combined effect of small bottleneck and vanishing view factor significantly lowers $\kappa_{eff}$. In the next section, we will analyze in detail the correlation between the pores arrangement and $\kappa_{eff}$.

\section*{Identification of a Descriptor}
 In previous work~\cite{Romano2014qj}, we reported that $\kappa_{eff}$ in nanoporous materials is dictated by the view factor and the pore-pore distances. We note that the view factor is a geometrical feature that describes the ability of a ray to travel across the simulation domain without intercepting the pores~\cite{howell2010thermal}. In random-pore configurations, the view factor vanishes because of the disordered pores blocking all the direct paths. It is natural, therefore, to speculate whether the average pore-pore distance in the disordered configurations is correlated with $\kappa_{eff}$. However, after a regression analysis, we conclude that unlike for the ordered case, such a parameter has only a marginal role for the disordered systems. In fact, rigorously speaking, only the interpore spaces perpendicular to heat flux matter. In order to identify the phonon bottleneck we then analyze the pores configuration in terms of graphs.

\begin{figure}[!ht]

\subfloat[\label{Fig:20a}]
{\includegraphics[width=0.48\columnwidth]{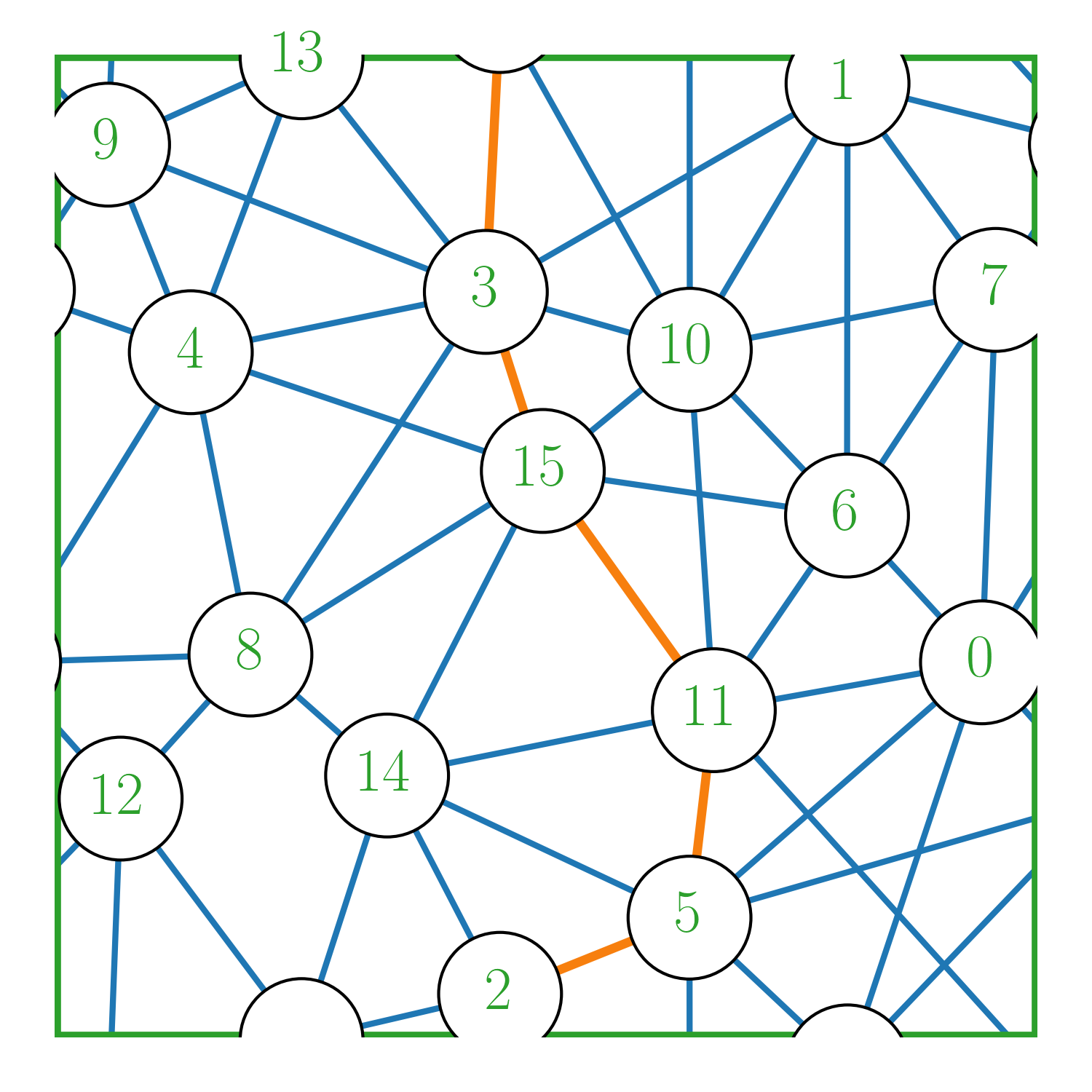}}
\hfill
\subfloat[\label{Fig:20b}]
{\includegraphics[width=0.48\columnwidth]{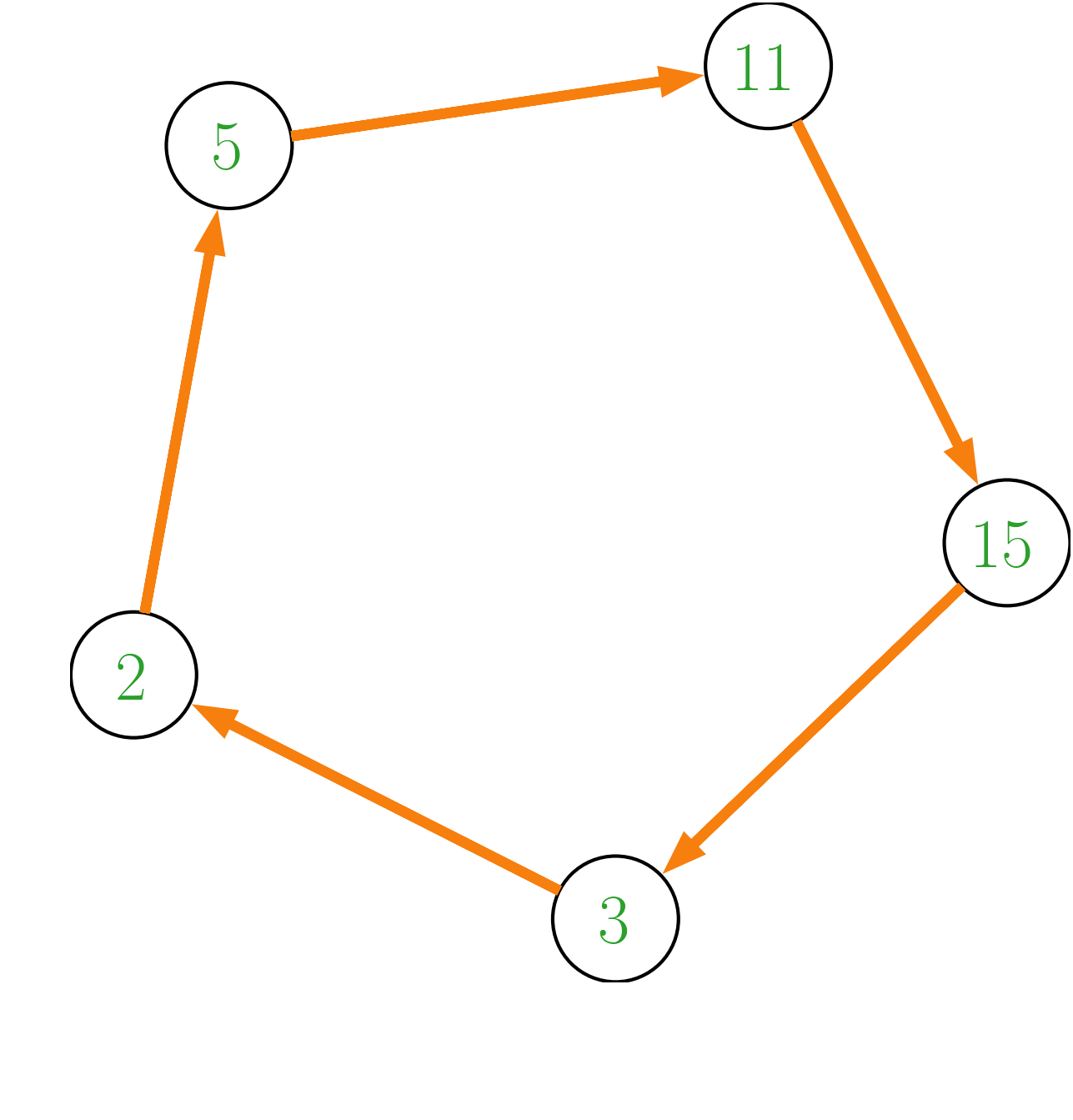}}
\caption{(a) Example of a first-neighbor map. Each pore in the unit cell is uniquely labeled. The bottleneck is highlighted by the orange line and, in (b), is represented by an elementary circuit, or cycle.}
\end{figure}

To this end, we first compute the pores first-neighbor map, as elaborated in the following. A given periodic configuration has a finite set of pores $P=\{P_0,P_1,...\,P_{N-1}\}$, where $N$ is the number of pores. Given two pores $P_{\alpha}$ and $P_{\beta}$, we define them to be neighbors if, when moving $P_{\alpha}$ toward $P_{\beta}$, there is no collision with the surrounding pores. The intersection among polygons is computed by the package PyClipper~\cite{pyclipper}. After repeating this procedure for all pore pairs, we obtain a first-neighbor map as shown in Fig.~\subref*{Fig:20a}. We then build the set of edges $E=\{E_0,E_1,...\,E_{M-1}\}$, where $M$ is the number of edges. Each edge connects two neighbor pores, say $P_\alpha$ and $P_{\beta}$, and points toward increasing $y$ coordinates, i.e., $P_{\alpha}$ is connected to $P_{\beta}$ only if $\left( \mathbf{C}_\beta - \mathbf{C}_\alpha\right)\cdot\mathbf{y} > 0 $, where $\mathbf{C}_n$ is the circumcenter of the pore $P_n$. The resulting graph, $G(P,E)$, is \textit{directed} in that its edges are unidirectional. We define a path in $G(P,E)$ as a sequence of vertices $p_{\mu \nu}=\{v_0= \mu,v_1,...\,v_{K-1}=\nu\}$ such that $\{v_i,v_{i+1}\}\in E$ for $0\le i<K-1$, where $K$ is the length of the path. An elementary circuit is a path where the only repeating vertexes are the first and the last ones, i.e, $\mu=\nu$. In a complete directed graph, the number of distinct elementary circuits, simply referred to as cycles, is \begin{equation}
\begin{split}\label{Eq:3}
S=\sum_{i=1}^{N-1}\binom{N}{N-i+1}(N-i)!,
\end{split}
\end{equation}which grows faster than $2^N$. Although in our case $G(P,E)$ is not complete, the number of cycles can easily reach a few thousand. Here we identify all possible cycles by using Johnson's algorithm, which has a time bound $O((N+M)(C+1))$~\cite{johnson1975finding}, where $C$ is the number of cycles. As the pores are identified uniquely within the unit cell, every pore shares the same label with its periodic counterpart. Consequently, the first and last nodes of a cycle, although having the same identifier, belong to two different unit cells. For our purposes, we select only cycles whose extreme nodes share the same $y$ coordinate, as exemplified in Fig.~\subref*{Fig:20b}. By doing so, we guarantee that the cycles are perpendicular to heat flow and, therefore, are suitable for the identification of a descriptor of thermal conductivity, as explained in the next section.

To identify the bottleneck for each configuration we develop the following algorithm:

\begin{enumerate}\item For each cycle, $\{C\}=C_0, C_s ... C_{S-1}$, we compute the interpore distances of its constituting pores, $\{R\}=R_0,R_k ... R_{K-1}$. Then, we compute the sum of such distances, i.e., $D_s=\sum_k R_k$. \item From the previous point, we have the set $\{D\}=D_0, D_s ... D_{S-1}$. The bottleneck is then $g = \mathrm{min}\{D\}$. \end{enumerate}

\begin{figure*}

\subfloat[\label{Fig:30a}]
{\includegraphics[width=0.48\textwidth]{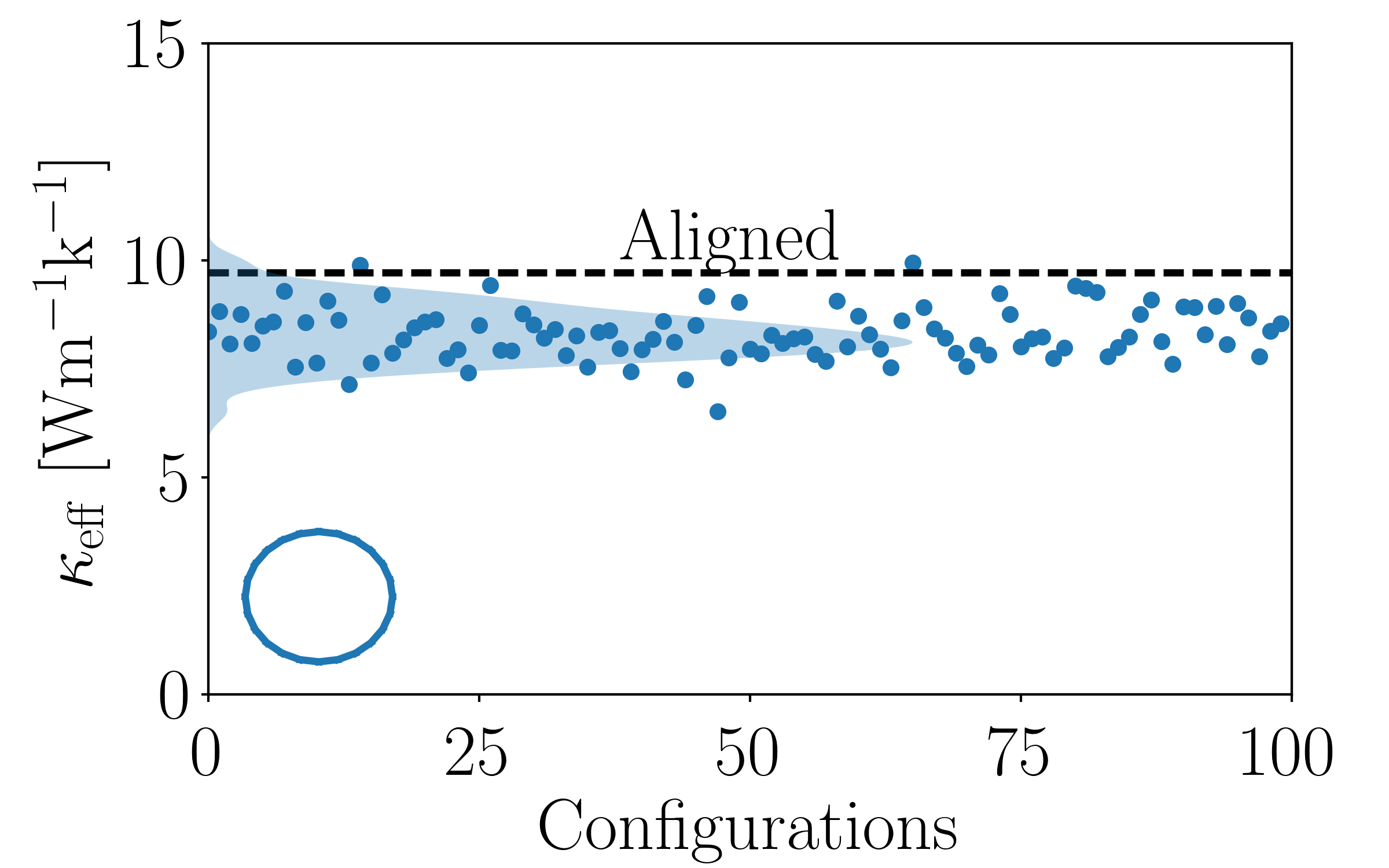}}
\hfill
\subfloat[\label{Fig:30b}]
{\includegraphics[width=0.48\textwidth]{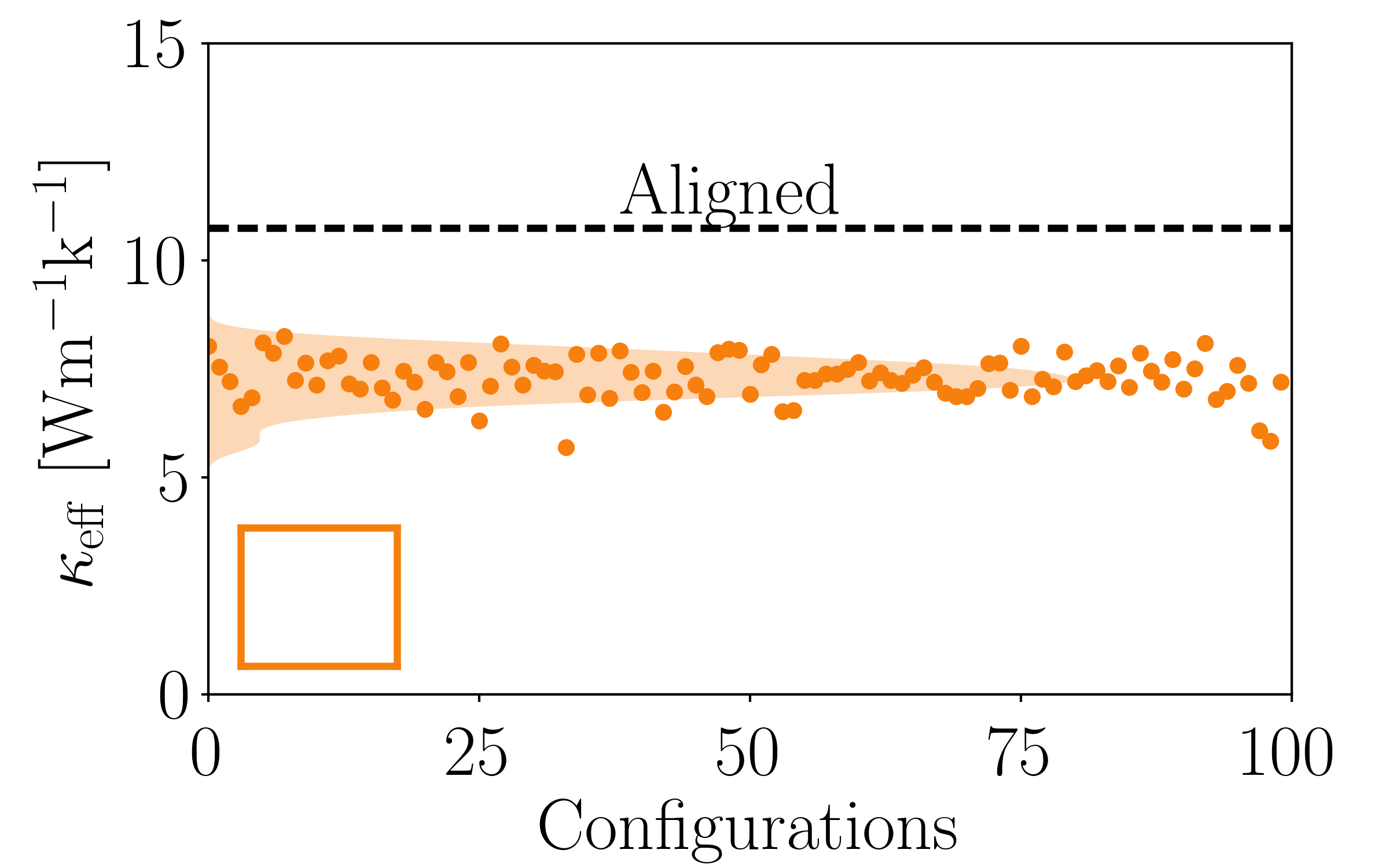}}
\quad
\subfloat[\label{Fig:30c}]
{\includegraphics[width=0.48\textwidth]{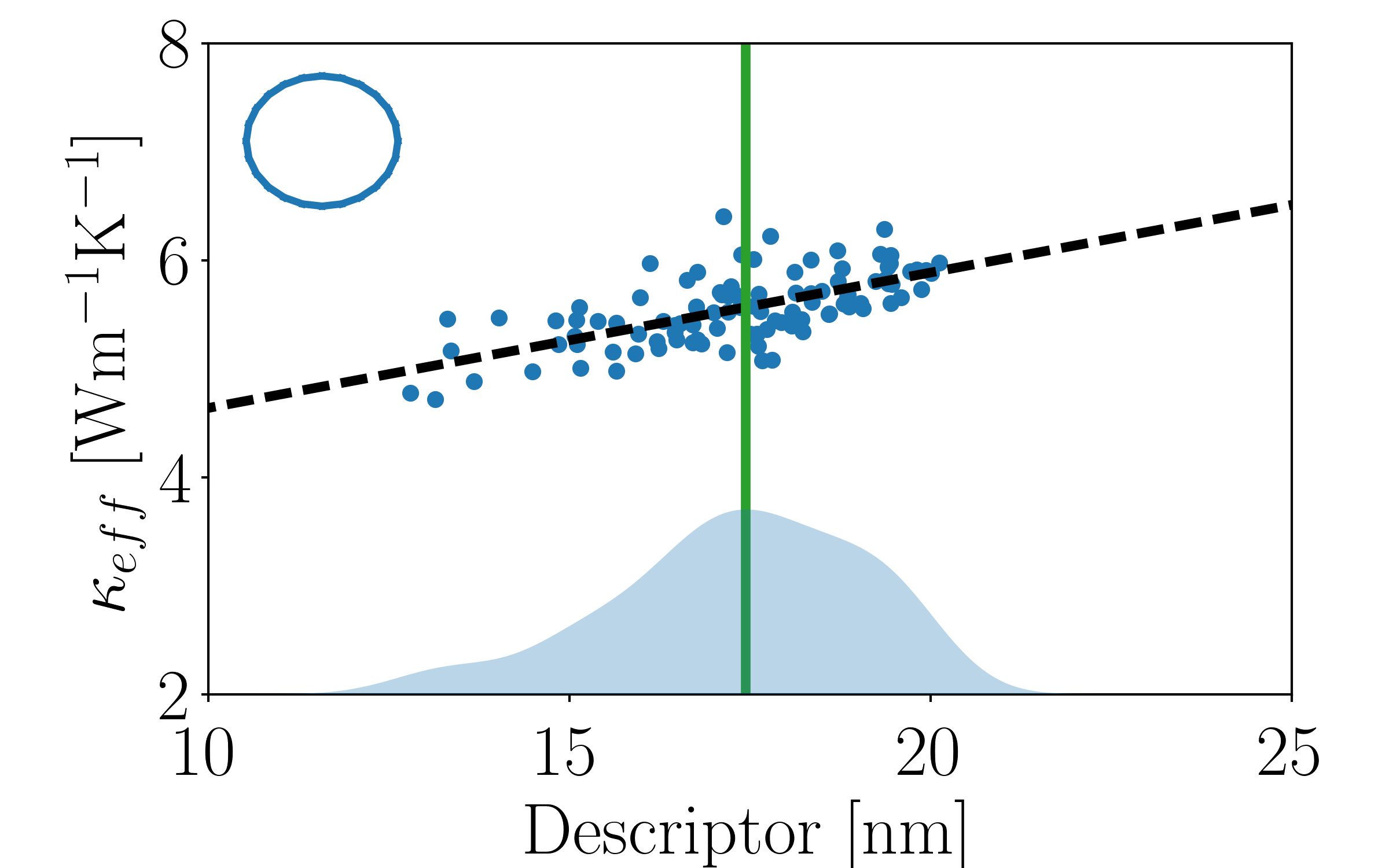}}
\hfill
\subfloat[\label{Fig:30d}]
{\includegraphics[width=0.48\textwidth]{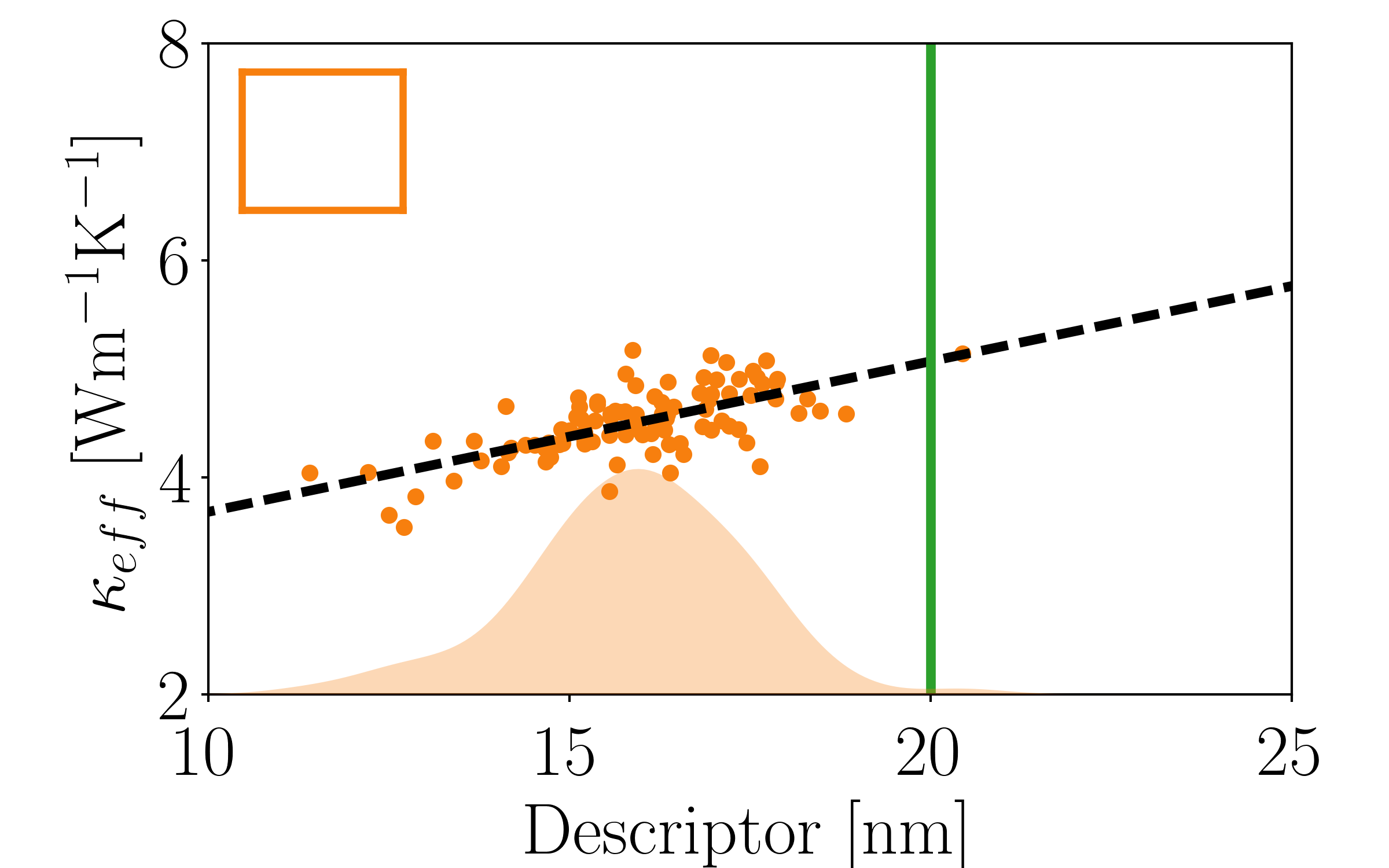}}
\caption{Distribution of $\kappa_{eff}$ for the (a) DC and (b) DS cases. The straight, horizontal lines represent the aligned counterparts. Distribution of $g$ for the (c) DC and (d) DS cases. The vertical lines refer to the bottleneck for the aligned cases.}
\end{figure*}
The phonon bottleneck is the smallest of the sum of pore-pore distances among all the cycles in a configuration. The effectiveness of $g$ in describing nanoscale thermal transport in such structures can be estimated by the Spearman correlation rank ($r_s$), a quantity that indicates how two variables are monotonically correlated to each other ~\cite{gastwirth1977lehmann}. The first step in computing $r_s$ is ranking the values for $\kappa_{eff}$ and $g$ and collecting the result via the vectors $\mathbf{K}$ and $\mathbf{G}$, respectively. Then, we compute \begin{equation}
\begin{split}\label{Eq:4}
r_s = 1-\frac{6\sum_i^n \left(G_i - K_i\right)}{n(n^2-1)},
\end{split}
\end{equation}where $n$ = 100 is the number of simulations for each shape. For both DC and DS cases, we obtain a significant Spearman correlation (higher than 0.63), suggesting that $g$ can be used as a good \textit{descriptor}. We use this knowledge to understand the $\kappa_{eff}$ distributions for the DC and DS cases in relation to the aligned cases. According to simple geometric considerations, the bottlneck for the aligned cases is simply $g_{AC} = 4L(1-2\sqrt{\phi/\pi})=$ 17.44 nm and $g_{AS}= 4L(1-\sqrt{\phi})$ = 20 nm. As shown in Fig.~\subref*{Fig:30c}, for DC, the average $g$ is around $g_{AC}$; for DS, almost all the configurations have $g$ smaller than that of AS [as shown in Fig.~\subref*{Fig:30d}], due to the square edges. These results reflect the relative trend in $\kappa_{eff}$ between the aligned and disordered cases, corroborating the use of $g$ as a valid descriptor for thermal transport. Moreover, we note that most of the bottlenecks have a number of pores ($\sim$6-7) higher than that of their aligned counterparts (4). This result confirms that smaller $\kappa_{eff}$, within configurations with the same porosity, can be achieved with anisotropic pore lattices, where the density of pores is higher along the Cartesian direction orthogonal to the applied temperature gradient~\cite{wolf2014monte,wolf2014thermal}. The introduction of a simple descriptor can be used to estimate the ranking of $\kappa_{eff}$ among different samples with disordered pores, supporting experiments on realistic materials~\cite{smith2016catalyst,Tang2010}. 
\section*{Conclusion}
In summary, by performing calculations of thermal transport in disordered porous materials we have quantified the effect of the randomness in pore arrangement on the thermal conductivity. Furthermore, we have devised a method to identify the set of special pores composing the phonon bottleneck, potentially empowering experimentalists with a simple tool to assess thermal conductivity in disordered porous materials.  

\begin{thebibliography}{31}%
\makeatletter
\providecommand \@ifxundefined [1]{%
 \@ifx{#1\undefined}
}%
\providecommand \@ifnum [1]{%
 \ifnum #1\expandafter \@firstoftwo
 \else \expandafter \@secondoftwo
 \fi
}%
\providecommand \@ifx [1]{%
 \ifx #1\expandafter \@firstoftwo
 \else \expandafter \@secondoftwo
 \fi
}%
\providecommand \natexlab [1]{#1}%
\providecommand \enquote  [1]{``#1''}%
\providecommand \bibnamefont  [1]{#1}%
\providecommand \bibfnamefont [1]{#1}%
\providecommand \citenamefont [1]{#1}%
\providecommand \href@noop [0]{\@secondoftwo}%
\providecommand \href [0]{\begingroup \@sanitize@url \@href}%
\providecommand \@href[1]{\@@startlink{#1}\@@href}%
\providecommand \@@href[1]{\endgroup#1\@@endlink}%
\providecommand \@sanitize@url [0]{\catcode `\\12\catcode `\$12\catcode
  `\&12\catcode `\#12\catcode `\^12\catcode `\_12\catcode `\%12\relax}%
\providecommand \@@startlink[1]{}%
\providecommand \@@endlink[0]{}%
\providecommand \url  [0]{\begingroup\@sanitize@url \@url }%
\providecommand \@url [1]{\endgroup\@href {#1}{\urlprefix }}%
\providecommand \urlprefix  [0]{URL }%
\providecommand \Eprint [0]{\href }%
\providecommand \doibase [0]{http://dx.doi.org/}%
\providecommand \selectlanguage [0]{\@gobble}%
\providecommand \bibinfo  [0]{\@secondoftwo}%
\providecommand \bibfield  [0]{\@secondoftwo}%
\providecommand \translation [1]{[#1]}%
\providecommand \BibitemOpen [0]{}%
\providecommand \bibitemStop [0]{}%
\providecommand \bibitemNoStop [0]{.\EOS\space}%
\providecommand \EOS [0]{\spacefactor3000\relax}%
\providecommand \BibitemShut  [1]{\csname bibitem#1\endcsname}%
\let\auto@bib@innerbib\@empty
\bibitem [{\citenamefont {Rowe}(1995)}]{CRC1995}%
  \BibitemOpen
  \bibinfo {editor} {\bibfnamefont {D.~M.}\ \bibnamefont {Rowe}},\ ed.,\ \href
  {http://dx.doi.org/10.1201/9781420049718} {\emph {\bibinfo {title} {{CRC}
  Handbook of Thermoelectrics}}}\ (\bibinfo  {publisher} {CRC Press, Boca
  Raton, FL},\ \bibinfo {year} {1995})\BibitemShut {NoStop}%
\bibitem [{\citenamefont {Tian}\ \emph {et~al.}(2013)\citenamefont {Tian},
  \citenamefont {Lee},\ and\ \citenamefont {Chen}}]{tian2013heat}%
  \BibitemOpen
  \bibfield  {author} {\bibinfo {author} {\bibfnamefont {Z.}~\bibnamefont
  {Tian}}, \bibinfo {author} {\bibfnamefont {S.}~\bibnamefont {Lee}}, \ and\
  \bibinfo {author} {\bibfnamefont {G.}~\bibnamefont {Chen}},\ }\href
  {http://heattransfer.asmedigitalcollection.asme.org/article.aspx?articleid=1688857}
  {\bibfield  {journal} {\bibinfo  {journal} {J. Heat Transfer}\ }\textbf
  {\bibinfo {volume} {135}},\ \bibinfo {pages} {061605} (\bibinfo {year}
  {2013})}\BibitemShut {NoStop}%
\bibitem [{\citenamefont {Liao}\ \emph {et~al.}(2015)\citenamefont {Liao},
  \citenamefont {Qiu}, \citenamefont {Zhou}, \citenamefont {Huberman},
  \citenamefont {Esfarjani},\ and\ \citenamefont {Chen}}]{liao2015significant}%
  \BibitemOpen
  \bibfield  {author} {\bibinfo {author} {\bibfnamefont {B.}~\bibnamefont
  {Liao}}, \bibinfo {author} {\bibfnamefont {B.}~\bibnamefont {Qiu}}, \bibinfo
  {author} {\bibfnamefont {J.}~\bibnamefont {Zhou}}, \bibinfo {author}
  {\bibfnamefont {S.}~\bibnamefont {Huberman}}, \bibinfo {author}
  {\bibfnamefont {K.}~\bibnamefont {Esfarjani}}, \ and\ \bibinfo {author}
  {\bibfnamefont {G.}~\bibnamefont {Chen}},\ }\href
  {link.aps.org/doi/10.1103/PhysRevLett.114.115901} {\bibfield  {journal}
  {\bibinfo  {journal} {Phys. Rev. Lett.}\ }\textbf {\bibinfo {volume} {114}},\
  \bibinfo {pages} {115901} (\bibinfo {year} {2015})}\BibitemShut {NoStop}%
\bibitem [{\citenamefont {Esfarjani}\ \emph {et~al.}(2011)\citenamefont
  {Esfarjani}, \citenamefont {Chen},\ and\ \citenamefont
  {Stokes}}]{esfarjani2011heat}%
  \BibitemOpen
  \bibfield  {author} {\bibinfo {author} {\bibfnamefont {K.}~\bibnamefont
  {Esfarjani}}, \bibinfo {author} {\bibfnamefont {G.}~\bibnamefont {Chen}}, \
  and\ \bibinfo {author} {\bibfnamefont {H.~T.}\ \bibnamefont {Stokes}},\
  }\href {link.aps.org/doi/10.1103/PhysRevB.84.085204} {\bibfield  {journal}
  {\bibinfo  {journal} {Phys. Rev. B}\ }\textbf {\bibinfo {volume} {84}},\
  \bibinfo {pages} {085204} (\bibinfo {year} {2011})}\BibitemShut {NoStop}%
\bibitem [{\citenamefont {Hopkins}\ \emph {et~al.}(2011)\citenamefont
  {Hopkins}, \citenamefont {Reinke}, \citenamefont {Su}, \citenamefont
  {Olsson}, \citenamefont {Shaner}, \citenamefont {Leseman}, \citenamefont
  {Serrano}, \citenamefont {Phinney},\ and\ \citenamefont
  {El-Kady}}]{Hopkins2011}%
  \BibitemOpen
  \bibfield  {author} {\bibinfo {author} {\bibfnamefont {P.~E.}\ \bibnamefont
  {Hopkins}}, \bibinfo {author} {\bibfnamefont {C.~M.}\ \bibnamefont {Reinke}},
  \bibinfo {author} {\bibfnamefont {M.~F.}\ \bibnamefont {Su}}, \bibinfo
  {author} {\bibfnamefont {R.~H.}\ \bibnamefont {Olsson}}, \bibinfo {author}
  {\bibfnamefont {E.~A.}\ \bibnamefont {Shaner}}, \bibinfo {author}
  {\bibfnamefont {Z.~C.}\ \bibnamefont {Leseman}}, \bibinfo {author}
  {\bibfnamefont {J.~R.}\ \bibnamefont {Serrano}}, \bibinfo {author}
  {\bibfnamefont {L.~M.}\ \bibnamefont {Phinney}}, \ and\ \bibinfo {author}
  {\bibfnamefont {I.}~\bibnamefont {El-Kady}},\ }\href
  {http://dx.doi.org/10.1021/nl102918q} {\bibfield  {journal} {\bibinfo
  {journal} {Nano Lett.}\ }\textbf {\bibinfo {volume} {11}},\ \bibinfo {pages}
  {107} (\bibinfo {year} {2011})}\BibitemShut {NoStop}%
\bibitem [{\citenamefont {Song}\ and\ \citenamefont
  {Chen}(2004)}]{song2004thermal}%
  \BibitemOpen
  \bibfield  {author} {\bibinfo {author} {\bibfnamefont {D.}~\bibnamefont
  {Song}}\ and\ \bibinfo {author} {\bibfnamefont {G.}~\bibnamefont {Chen}},\
  }\href {http://dx.doi.org/10.1063/1.1642753} {\bibfield  {journal} {\bibinfo
  {journal} {Appl. Phys. Lett.}\ }\textbf {\bibinfo {volume} {84}},\ \bibinfo
  {pages} {687} (\bibinfo {year} {2004})}\BibitemShut {NoStop}%
\bibitem [{\citenamefont {Lee}\ \emph {et~al.}(2015)\citenamefont {Lee},
  \citenamefont {Lim},\ and\ \citenamefont {Yang}}]{lee2015ballistic}%
  \BibitemOpen
  \bibfield  {author} {\bibinfo {author} {\bibfnamefont {J.}~\bibnamefont
  {Lee}}, \bibinfo {author} {\bibfnamefont {J.}~\bibnamefont {Lim}}, \ and\
  \bibinfo {author} {\bibfnamefont {P.}~\bibnamefont {Yang}},\ }\href
  {http://pubs.acs.org/doi/abs/10.1021/acs.nanolett.5b00495} {\bibfield
  {journal} {\bibinfo  {journal} {Nano Lett.}\ }\textbf {\bibinfo {volume}
  {15}},\ \bibinfo {pages} {3273} (\bibinfo {year} {2015})}\BibitemShut
  {NoStop}%
\bibitem [{\citenamefont {Tang}\ \emph {et~al.}(2010)\citenamefont {Tang},
  \citenamefont {Wang}, \citenamefont {Lee}, \citenamefont {Fardy},
  \citenamefont {Huo}, \citenamefont {Russell},\ and\ \citenamefont
  {Yang}}]{Tang2010}%
  \BibitemOpen
  \bibfield  {author} {\bibinfo {author} {\bibfnamefont {J.}~\bibnamefont
  {Tang}}, \bibinfo {author} {\bibfnamefont {H.-T.}\ \bibnamefont {Wang}},
  \bibinfo {author} {\bibfnamefont {D.~H.}\ \bibnamefont {Lee}}, \bibinfo
  {author} {\bibfnamefont {M.}~\bibnamefont {Fardy}}, \bibinfo {author}
  {\bibfnamefont {Z.}~\bibnamefont {Huo}}, \bibinfo {author} {\bibfnamefont
  {T.~P.}\ \bibnamefont {Russell}}, \ and\ \bibinfo {author} {\bibfnamefont
  {P.}~\bibnamefont {Yang}},\ }\href {http://dx.doi.org/10.1021/nl102931z}
  {\bibfield  {journal} {\bibinfo  {journal} {Nano Lett.}\ }\textbf {\bibinfo
  {volume} {10}},\ \bibinfo {pages} {4279} (\bibinfo {year}
  {2010})}\BibitemShut {NoStop}%
\bibitem [{\citenamefont {Yu}\ \emph {et~al.}(2010)\citenamefont {Yu},
  \citenamefont {Mitrovic}, \citenamefont {Tham}, \citenamefont {Varghese},\
  and\ \citenamefont {Heath}}]{Yu2010}%
  \BibitemOpen
  \bibfield  {author} {\bibinfo {author} {\bibfnamefont {J.-K.}\ \bibnamefont
  {Yu}}, \bibinfo {author} {\bibfnamefont {S.}~\bibnamefont {Mitrovic}},
  \bibinfo {author} {\bibfnamefont {D.}~\bibnamefont {Tham}}, \bibinfo {author}
  {\bibfnamefont {J.}~\bibnamefont {Varghese}}, \ and\ \bibinfo {author}
  {\bibfnamefont {J.~R.}\ \bibnamefont {Heath}},\ }\href
  {http://dx.doi.org/10.1038/nnano.2010.149} {\bibfield  {journal} {\bibinfo
  {journal} {Nat. Nanotechnol.}\ }\textbf {\bibinfo {volume} {5}},\ \bibinfo
  {pages} {718} (\bibinfo {year} {2010})}\BibitemShut {NoStop}%
\bibitem [{\citenamefont {Verdier}\ \emph {et~al.}(2017)\citenamefont
  {Verdier}, \citenamefont {Anufriev}, \citenamefont {Ramiere}, \citenamefont
  {Termentzidis},\ and\ \citenamefont {Lacroix}}]{verdier2017thermal}%
  \BibitemOpen
  \bibfield  {author} {\bibinfo {author} {\bibfnamefont {M.}~\bibnamefont
  {Verdier}}, \bibinfo {author} {\bibfnamefont {R.}~\bibnamefont {Anufriev}},
  \bibinfo {author} {\bibfnamefont {A.}~\bibnamefont {Ramiere}}, \bibinfo
  {author} {\bibfnamefont {K.}~\bibnamefont {Termentzidis}}, \ and\ \bibinfo
  {author} {\bibfnamefont {D.}~\bibnamefont {Lacroix}},\ }\href
  {https://journals.aps.org/prb/abstract/10.1103/PhysRevB.95.205438} {\bibfield
   {journal} {\bibinfo  {journal} {Physical Review B}\ }\textbf {\bibinfo
  {volume} {95}},\ \bibinfo {pages} {205438} (\bibinfo {year}
  {2017})}\BibitemShut {NoStop}%
\bibitem [{\citenamefont {Perez-Taborda}\ \emph {et~al.}(2016)\citenamefont
  {Perez-Taborda}, \citenamefont {Rojo}, \citenamefont {Maiz}, \citenamefont
  {Neophytou},\ and\ \citenamefont {Martin-Gonzalez}}]{perez2016ultra}%
  \BibitemOpen
  \bibfield  {author} {\bibinfo {author} {\bibfnamefont {J.~A.}\ \bibnamefont
  {Perez-Taborda}}, \bibinfo {author} {\bibfnamefont {M.~M.}\ \bibnamefont
  {Rojo}}, \bibinfo {author} {\bibfnamefont {J.}~\bibnamefont {Maiz}}, \bibinfo
  {author} {\bibfnamefont {N.}~\bibnamefont {Neophytou}}, \ and\ \bibinfo
  {author} {\bibfnamefont {M.}~\bibnamefont {Martin-Gonzalez}},\ }\href
  {https://www.nature.com/articles/srep32778} {\bibfield  {journal} {\bibinfo
  {journal} {Scientific reports}\ }\textbf {\bibinfo {volume} {6}},\ \bibinfo
  {pages} {32778} (\bibinfo {year} {2016})}\BibitemShut {NoStop}%
\bibitem [{\citenamefont {Lee}\ \emph {et~al.}(2017)\citenamefont {Lee},
  \citenamefont {Lee}, \citenamefont {Wehmeyer}, \citenamefont {Dhuey},
  \citenamefont {Olynick}, \citenamefont {Cabrini}, \citenamefont {Dames},
  \citenamefont {Urban},\ and\ \citenamefont {Yang}}]{lee2017investigation}%
  \BibitemOpen
  \bibfield  {author} {\bibinfo {author} {\bibfnamefont {J.}~\bibnamefont
  {Lee}}, \bibinfo {author} {\bibfnamefont {W.}~\bibnamefont {Lee}}, \bibinfo
  {author} {\bibfnamefont {G.}~\bibnamefont {Wehmeyer}}, \bibinfo {author}
  {\bibfnamefont {S.}~\bibnamefont {Dhuey}}, \bibinfo {author} {\bibfnamefont
  {D.~L.}\ \bibnamefont {Olynick}}, \bibinfo {author} {\bibfnamefont
  {S.}~\bibnamefont {Cabrini}}, \bibinfo {author} {\bibfnamefont
  {C.}~\bibnamefont {Dames}}, \bibinfo {author} {\bibfnamefont {J.~J.}\
  \bibnamefont {Urban}}, \ and\ \bibinfo {author} {\bibfnamefont
  {P.}~\bibnamefont {Yang}},\ }\href
  {https://www.nature.com/articles/ncomms14054?WT.feed_name=subjects_physics}
  {\bibfield  {journal} {\bibinfo  {journal} {Nature communications}\ }\textbf
  {\bibinfo {volume} {8}},\ \bibinfo {pages} {14054} (\bibinfo {year}
  {2017})}\BibitemShut {NoStop}%
\bibitem [{\citenamefont {Hao}\ \emph {et~al.}(2009)\citenamefont {Hao},
  \citenamefont {Chen},\ and\ \citenamefont {Jeng}}]{hao2009frequency}%
  \BibitemOpen
  \bibfield  {author} {\bibinfo {author} {\bibfnamefont {Q.}~\bibnamefont
  {Hao}}, \bibinfo {author} {\bibfnamefont {G.}~\bibnamefont {Chen}}, \ and\
  \bibinfo {author} {\bibfnamefont {M.-S.}\ \bibnamefont {Jeng}},\ }\href
  {http://dx.doi.org/10.1063/1.3266169} {\bibfield  {journal} {\bibinfo
  {journal} {J. Appl. Phys.}\ }\textbf {\bibinfo {volume} {106}},\ \bibinfo
  {pages} {114321} (\bibinfo {year} {2009})}\BibitemShut {NoStop}%
\bibitem [{\citenamefont {Lee}\ \emph {et~al.}(2008)\citenamefont {Lee},
  \citenamefont {Galli},\ and\ \citenamefont {Grossman}}]{lee2008nanoporous}%
  \BibitemOpen
  \bibfield  {author} {\bibinfo {author} {\bibfnamefont {J.-H.}\ \bibnamefont
  {Lee}}, \bibinfo {author} {\bibfnamefont {G.~A.}\ \bibnamefont {Galli}}, \
  and\ \bibinfo {author} {\bibfnamefont {J.~C.}\ \bibnamefont {Grossman}},\
  }\href {http://pubs.acs.org/doi/abs/10.1021/nl802045} {\bibfield  {journal}
  {\bibinfo  {journal} {Nano Lett.}\ }\textbf {\bibinfo {volume} {8}},\
  \bibinfo {pages} {3750} (\bibinfo {year} {2008})}\BibitemShut {NoStop}%
\bibitem [{\citenamefont {Romano}\ and\ \citenamefont
  {Grossman}(2015)}]{romano2015}%
  \BibitemOpen
  \bibfield  {author} {\bibinfo {author} {\bibfnamefont {G.}~\bibnamefont
  {Romano}}\ and\ \bibinfo {author} {\bibfnamefont {J.~C.}\ \bibnamefont
  {Grossman}},\ }\href
  {https://heattransfer.asmedigitalcollection.asme.org/article.aspx?articleid=2119334}
  {\bibfield  {journal} {\bibinfo  {journal} {J. Heat Transf.}\ }\textbf
  {\bibinfo {volume} {137}},\ \bibinfo {pages} {071302} (\bibinfo {year}
  {2015})}\BibitemShut {NoStop}%
\bibitem [{\citenamefont {Romano}\ and\ \citenamefont
  {Grossman}(2014)}]{Romano2014qj}%
  \BibitemOpen
  \bibfield  {author} {\bibinfo {author} {\bibfnamefont {G.}~\bibnamefont
  {Romano}}\ and\ \bibinfo {author} {\bibfnamefont {J.~C.}\ \bibnamefont
  {Grossman}},\ }\href {http://aip.scitation.org/doi/full/10.1063/1.4891362}
  {\bibfield  {journal} {\bibinfo  {journal} {Appl. Phys. Lett.}\ }\textbf
  {\bibinfo {volume} {105}},\ \bibinfo {pages} {033116} (\bibinfo {year}
  {2014})}\BibitemShut {NoStop}%
\bibitem [{\citenamefont {Smith}\ \emph {et~al.}(2016)\citenamefont {Smith},
  \citenamefont {Patil}, \citenamefont {Ferralis},\ and\ \citenamefont
  {Grossman}}]{smith2016catalyst}%
  \BibitemOpen
  \bibfield  {author} {\bibinfo {author} {\bibfnamefont {B.~D.}\ \bibnamefont
  {Smith}}, \bibinfo {author} {\bibfnamefont {J.~J.}\ \bibnamefont {Patil}},
  \bibinfo {author} {\bibfnamefont {N.}~\bibnamefont {Ferralis}}, \ and\
  \bibinfo {author} {\bibfnamefont {J.~C.}\ \bibnamefont {Grossman}},\ }\href
  {pubs.acs.org/doi/abs/10.1021/acsami.6b01927} {\bibfield  {journal} {\bibinfo
   {journal} {ACS Appl. Mater. Interfaces}\ }\textbf {\bibinfo {volume} {8}},\
  \bibinfo {pages} {8043} (\bibinfo {year} {2016})}\BibitemShut {NoStop}%
\bibitem [{\citenamefont {Wolf}\ \emph
  {et~al.}(2014{\natexlab{a}})\citenamefont {Wolf}, \citenamefont {Neophytou},
  \citenamefont {Stanojevic},\ and\ \citenamefont {Kosina}}]{wolf2014monte}%
  \BibitemOpen
  \bibfield  {author} {\bibinfo {author} {\bibfnamefont {S.}~\bibnamefont
  {Wolf}}, \bibinfo {author} {\bibfnamefont {N.}~\bibnamefont {Neophytou}},
  \bibinfo {author} {\bibfnamefont {Z.}~\bibnamefont {Stanojevic}}, \ and\
  \bibinfo {author} {\bibfnamefont {H.}~\bibnamefont {Kosina}},\ }\href
  {http://link.springer.com/article/10.1007/s11664-014-3324-x} {\bibfield
  {journal} {\bibinfo  {journal} {J. Electron. Mater.}\ }\textbf {\bibinfo
  {volume} {43}},\ \bibinfo {pages} {3870} (\bibinfo {year}
  {2014}{\natexlab{a}})}\BibitemShut {NoStop}%
\bibitem [{\citenamefont {Wolf}\ \emph
  {et~al.}(2014{\natexlab{b}})\citenamefont {Wolf}, \citenamefont {Neophytou},\
  and\ \citenamefont {Kosina}}]{wolf2014thermal}%
  \BibitemOpen
  \bibfield  {author} {\bibinfo {author} {\bibfnamefont {S.}~\bibnamefont
  {Wolf}}, \bibinfo {author} {\bibfnamefont {N.}~\bibnamefont {Neophytou}}, \
  and\ \bibinfo {author} {\bibfnamefont {H.}~\bibnamefont {Kosina}},\ }\href
  {http://aip.scitation.org/doi/abs/10.1063/1.4879242} {\bibfield  {journal}
  {\bibinfo  {journal} {J. Appl. Phys.}\ }\textbf {\bibinfo {volume} {115}},\
  \bibinfo {pages} {204306} (\bibinfo {year} {2014}{\natexlab{b}})}\BibitemShut
  {NoStop}%
\bibitem [{\citenamefont {Romano}\ \emph {et~al.}(2016)\citenamefont {Romano},
  \citenamefont {Esfarjani}, \citenamefont {Strubbe}, \citenamefont {Broido},\
  and\ \citenamefont {Kolpak}}]{romano2016temperature}%
  \BibitemOpen
  \bibfield  {author} {\bibinfo {author} {\bibfnamefont {G.}~\bibnamefont
  {Romano}}, \bibinfo {author} {\bibfnamefont {K.}~\bibnamefont {Esfarjani}},
  \bibinfo {author} {\bibfnamefont {D.~A.}\ \bibnamefont {Strubbe}}, \bibinfo
  {author} {\bibfnamefont {D.}~\bibnamefont {Broido}}, \ and\ \bibinfo {author}
  {\bibfnamefont {A.~M.}\ \bibnamefont {Kolpak}},\ }\href
  {https://doi.org/10.1103/PhysRevB.93.035408} {\bibfield  {journal} {\bibinfo
  {journal} {Phys. Rev. B}\ }\textbf {\bibinfo {volume} {93}},\ \bibinfo
  {pages} {035408} (\bibinfo {year} {2016})}\BibitemShut {NoStop}%
\bibitem [{\citenamefont {Broido}\ \emph {et~al.}(2007)\citenamefont {Broido},
  \citenamefont {Malorny}, \citenamefont {Birner}, \citenamefont {Mingo},\ and\
  \citenamefont {Stewart}}]{broido2007intrinsic}%
  \BibitemOpen
  \bibfield  {author} {\bibinfo {author} {\bibfnamefont {D.}~\bibnamefont
  {Broido}}, \bibinfo {author} {\bibfnamefont {M.}~\bibnamefont {Malorny}},
  \bibinfo {author} {\bibfnamefont {G.}~\bibnamefont {Birner}}, \bibinfo
  {author} {\bibfnamefont {N.}~\bibnamefont {Mingo}}, \ and\ \bibinfo {author}
  {\bibfnamefont {D.}~\bibnamefont {Stewart}},\ }\href
  {http://dx.doi.org/10.1063/1.2822891} {\bibfield  {journal} {\bibinfo
  {journal} {Appl. Phys. Lett.}\ }\textbf {\bibinfo {volume} {91}},\ \bibinfo
  {pages} {231922} (\bibinfo {year} {2007})}\BibitemShut {NoStop}%
\bibitem [{\citenamefont {Geuzaine}\ and\ \citenamefont
  {Remacle}(2009)}]{geuzaine2009gmsh}%
  \BibitemOpen
  \bibfield  {author} {\bibinfo {author} {\bibfnamefont {C.}~\bibnamefont
  {Geuzaine}}\ and\ \bibinfo {author} {\bibfnamefont {J.-F.}\ \bibnamefont
  {Remacle}},\ }\href
  {http://onlinelibrary.wiley.com/doi/10.1002/nme.2579/full} {\bibfield
  {journal} {\bibinfo  {journal} {Int. J. Numer. Meth. Eng.}\ }\textbf
  {\bibinfo {volume} {79}},\ \bibinfo {pages} {1309} (\bibinfo {year}
  {2009})}\BibitemShut {NoStop}%
\bibitem [{\citenamefont {Abe}(1997)}]{abe1997derivation}%
  \BibitemOpen
  \bibfield  {author} {\bibinfo {author} {\bibfnamefont {T.}~\bibnamefont
  {Abe}},\ }\href
  {http://www.sciencedirect.com/science/article/pii/S0021999196955953}
  {\bibfield  {journal} {\bibinfo  {journal} {J. Comput. Phys.}\ }\textbf
  {\bibinfo {volume} {131}},\ \bibinfo {pages} {241} (\bibinfo {year}
  {1997})}\BibitemShut {NoStop}%
\bibitem [{\citenamefont {Romano}\ and\ \citenamefont
  {Di~Carlo}(2011)}]{romano2011multiscale}%
  \BibitemOpen
  \bibfield  {author} {\bibinfo {author} {\bibfnamefont {G.}~\bibnamefont
  {Romano}}\ and\ \bibinfo {author} {\bibfnamefont {A.}~\bibnamefont
  {Di~Carlo}},\ }\href
  {http://ieeexplore.ieee.org/document/5740609/?arnumber=5740609&tag=1}
  {\bibfield  {journal} {\bibinfo  {journal} {IEEE Trans. Nanotechnol.}\
  }\textbf {\bibinfo {volume} {10}},\ \bibinfo {pages} {1285} (\bibinfo {year}
  {2011})}\BibitemShut {NoStop}%
\bibitem [{\citenamefont {Li}\ \emph {et~al.}(2014)\citenamefont {Li},
  \citenamefont {Carrete}, \citenamefont {A.~Katcho},\ and\ \citenamefont
  {Mingo}}]{Li2014fg}%
  \BibitemOpen
  \bibfield  {author} {\bibinfo {author} {\bibfnamefont {W.}~\bibnamefont
  {Li}}, \bibinfo {author} {\bibfnamefont {J.}~\bibnamefont {Carrete}},
  \bibinfo {author} {\bibfnamefont {N.}~\bibnamefont {A.~Katcho}}, \ and\
  \bibinfo {author} {\bibfnamefont {N.}~\bibnamefont {Mingo}},\ }\href
  {http://www.sciencedirect.com/science/article/pii/S0010465514000484}
  {\bibfield  {journal} {\bibinfo  {journal} {Comput. Phys. Commun.}\ }\textbf
  {\bibinfo {volume} {185}},\ \bibinfo {pages} {1747} (\bibinfo {year}
  {2014})}\BibitemShut {NoStop}%
\bibitem [{\citenamefont {Vega-Flick}\ \emph {et~al.}(2016)\citenamefont
  {Vega-Flick}, \citenamefont {Duncan}, \citenamefont {Eliason}, \citenamefont
  {Cuffe}, \citenamefont {Johnson}, \citenamefont {Peraud}, \citenamefont
  {Zeng}, \citenamefont {Lu}, \citenamefont {Maznev}, \citenamefont {Wang}
  \emph {et~al.}}]{vega2016thermal}%
  \BibitemOpen
  \bibfield  {author} {\bibinfo {author} {\bibfnamefont {A.}~\bibnamefont
  {Vega-Flick}}, \bibinfo {author} {\bibfnamefont {R.~A.}\ \bibnamefont
  {Duncan}}, \bibinfo {author} {\bibfnamefont {J.~K.}\ \bibnamefont {Eliason}},
  \bibinfo {author} {\bibfnamefont {J.}~\bibnamefont {Cuffe}}, \bibinfo
  {author} {\bibfnamefont {J.~A.}\ \bibnamefont {Johnson}}, \bibinfo {author}
  {\bibfnamefont {J.-P.}\ \bibnamefont {Peraud}}, \bibinfo {author}
  {\bibfnamefont {L.}~\bibnamefont {Zeng}}, \bibinfo {author} {\bibfnamefont
  {Z.}~\bibnamefont {Lu}}, \bibinfo {author} {\bibfnamefont {A.~A.}\
  \bibnamefont {Maznev}}, \bibinfo {author} {\bibfnamefont {E.~N.}\
  \bibnamefont {Wang}},  \emph {et~al.},\ }\href
  {http://aip.scitation.org/doi/full/10.1063/1.4968610} {\bibfield  {journal}
  {\bibinfo  {journal} {AIP Adv.}\ }\textbf {\bibinfo {volume} {6}},\ \bibinfo
  {pages} {121903} (\bibinfo {year} {2016})}\BibitemShut {NoStop}%
\bibitem [{\citenamefont {Park}\ \emph {et~al.}(2017)\citenamefont {Park},
  \citenamefont {Romano}, \citenamefont {Ahn}, \citenamefont {Kodama},
  \citenamefont {Park}, \citenamefont {Barako}, \citenamefont {Sohn J.and~Cho},
  \citenamefont {Kim}, \citenamefont {Marconnet}, \citenamefont {Asheghi},
  \citenamefont {Kolpak}, \citenamefont {Sinclair},\ and\ \citenamefont
  {Goodson}}]{goodson}%
  \BibitemOpen
  \bibfield  {author} {\bibinfo {author} {\bibfnamefont {W.}~\bibnamefont
  {Park}}, \bibinfo {author} {\bibfnamefont {G.}~\bibnamefont {Romano}},
  \bibinfo {author} {\bibfnamefont {E.~C.}\ \bibnamefont {Ahn}}, \bibinfo
  {author} {\bibfnamefont {T.}~\bibnamefont {Kodama}}, \bibinfo {author}
  {\bibfnamefont {J.}~\bibnamefont {Park}}, \bibinfo {author} {\bibfnamefont
  {M.~T.}\ \bibnamefont {Barako}}, \bibinfo {author} {\bibfnamefont
  {J.}~\bibnamefont {Sohn J.and~Cho}}, \bibinfo {author} {\bibfnamefont
  {S.~J.}\ \bibnamefont {Kim}}, \bibinfo {author} {\bibfnamefont {A.~M.}\
  \bibnamefont {Marconnet}}, \bibinfo {author} {\bibfnamefont {M.}~\bibnamefont
  {Asheghi}}, \bibinfo {author} {\bibfnamefont {A.~M.}\ \bibnamefont {Kolpak}},
  \bibinfo {author} {\bibfnamefont {R.}~\bibnamefont {Sinclair}}, \ and\
  \bibinfo {author} {\bibfnamefont {K.~E.}\ \bibnamefont {Goodson}},\ }\href
  {https://www.ncbi.nlm.nih.gov/pmc/articles/PMC5524879/} {\bibfield  {journal}
  {\bibinfo  {journal} {Sci. Rep.}\ }\textbf {\bibinfo {volume} {7}},\ \bibinfo
  {pages} {6233} (\bibinfo {year} {2017})}\BibitemShut {NoStop}%
\bibitem [{\citenamefont {Howell}\ \emph {et~al.}(2010)\citenamefont {Howell},
  \citenamefont {Menguc},\ and\ \citenamefont {Siegel}}]{howell2010thermal}%
  \BibitemOpen
  \bibfield  {author} {\bibinfo {author} {\bibfnamefont {J.~R.}\ \bibnamefont
  {Howell}}, \bibinfo {author} {\bibfnamefont {M.~P.}\ \bibnamefont {Menguc}},
  \ and\ \bibinfo {author} {\bibfnamefont {R.}~\bibnamefont {Siegel}},\ }\href
  {https://www.crcpress.com/Thermal-Radiation-Heat-Transfer-5th-Edition/Howell-Menguc-Siegel/p/book/9781439866689}
  {\emph {\bibinfo {title} {Thermal radiation heat transfer}}}\ (\bibinfo
  {publisher} {CRC press},\ \bibinfo {address} {Boca Raton, FL},\ \bibinfo
  {year} {2010})\BibitemShut {NoStop}%
\bibitem [{pyc()}]{pyclipper}%
  \BibitemOpen
  \href {https://pypi.python.org/pypi/pyclipper} {\bibinfo  {journal}
  {PyClipper 1.0.2. https://pypi.python.org/pypi/pyclipper}\ }\BibitemShut
  {NoStop}%
\bibitem [{\citenamefont {Johnson}(1975)}]{johnson1975finding}%
  \BibitemOpen
\bibfield  {journal} {  }\bibfield  {author} {\bibinfo {author} {\bibfnamefont
  {D.~B.}\ \bibnamefont {Johnson}},\ }\href
  {http://epubs.siam.org/doi/abs/10.1137/0204007} {\bibfield  {journal}
  {\bibinfo  {journal} {SIAM J. Comput.}\ }\textbf {\bibinfo {volume} {4}},\
  \bibinfo {pages} {77} (\bibinfo {year} {1975})}\BibitemShut {NoStop}%
\bibitem [{\citenamefont {Lehmann}\ and\ \citenamefont
  {D'abrera}(2006)}]{gastwirth1977lehmann}%
  \BibitemOpen
  \bibfield  {author} {\bibinfo {author} {\bibfnamefont {E.~L.}\ \bibnamefont
  {Lehmann}}\ and\ \bibinfo {author} {\bibfnamefont {H.}~\bibnamefont
  {D'abrera}},\ }\href {http://www.springer.com/us/book/9780387352121} {\emph
  {\bibinfo {title} {Nonparametrics: statistical methods based on ranks.}}}\
  (\bibinfo  {publisher} {Springer},\ \bibinfo {address} {New York City},\
  \bibinfo {year} {2006})\BibitemShut {NoStop}%
\end{thebibliography}
\end{document}